# Machine Ethics: The Creation of a Virtuous Machine

Mohamed Akrout [(1)] and Robert Steinbauer [(2)]

*(1) Department of Computer Science, University of Toronto, 214 College St, Toronto, ON M5T 3A1, Canada*
*(2) OBHREE, Brock University, 1812 Sir Isaac Brock Way, St. Catharines, ON L2S3A1, Canada*

Artificial intelligence (AI) was initially developed as an implicit moral agent to solve simple and clearly defined tasks where all options are predictable. However, it is now part of our daily life powering cell phones, cameras, watches, thermostats, vacuums, cars, and much more. This has raised numerous concerns and some scholars and practitioners stress the dangers of AI and argue against its development as moral agents that can reason about ethics (e.g., Bryson 2008; Johnson and Miller 2008; Sharkey 2017; Tonkens 2009; van Wynsberghe and Robbins 2019). Even though we acknowledge the potential threat, in line with most other scholars (e.g., Anderson and Anderson 2010; Moor 2006; Scheutz 2016; Wallach 2010), we argue that AI advancements cannot be stopped and developers need to prepare AI to sustain explicit moral agents and face ethical dilemmas in complex and morally salient environments.

Explicit moral agents need to independently engage in moral reasoning to find solutions in situations where two or more ethical imperatives collide. Ethical theories build the foundation for such moral reasoning and have been used in the past to support AI decision making. Anderson and Anderson (2007) for example applied a Utilitarian perspective and discussed how AI can be used to calculate the total net pleasure of an act. However, Utilitarianism is only beneficial when all options are identifiable and the outcomes can be assessed in terms of pleasure and displeasure. Consequently, AI scholars now seem to favour a deontological approach to machine ethics that stresses the importance of justice and principles (Kant 2013). However, current deontological approaches (e.g., Anderson, Anderson, and Berenz's (2018) eldercare robot) require that all ethical principles are predefined and applied using static rule-based filtering. The exhaustive search of ethical principles becomes intractable in complex cases such as self-driving cars and AI-based diagnosis systems. Even if the ethical principles that should be applied have been hard coded by AI engineers, the balance between these principles remains an open question.

Despite the limitations discussed, AI Ethics scholars still heavily focus on outcome and action-focused approaches and dismiss Virtue Ethics since it is agent-based (Anderson and Anderson 2007) (for an exception see Govindarajulu et al. 2019). This is unfortunate as ignoring knowledge from the domain of Virtue Ethics will potentially result in flawed explicit moral agents who narrowly focus on outcomes and actions and neglect to include ethical metrics into the recursive policy learning process that shapes future AI decisions. Hence, the purpose of this research is to review the virtue ethics literature, identify important characteristics, and apply them to the AI context.

Virtue Ethics focuses on the virtues a decision maker should possess. Confucius, Plato, Aristotle, Hume, and Nietzsche have developed slightly different forms of virtue ethics (Swanton 2003). However, at the core of each approach are virtues that represent deeply integrated character traits. One should not single out a specific virtue, but strive for the golden mean, a state where virtues are balanced. For example, Aristotle focused on justice, charity, courage, truthfulness, modesty, and friendliness while Confucius stressed patience, knowledge, sincerity, and obedience. To be virtuous means that one tells the truth not because of the consequences of lying or because lying is bad, but because of his or her views about honesty and deception. Virtue Ethics is agent based and focuses on the person by raising questions such as: What kind of person should I be? To be a virtuous person, an individual has to have the right intentions. While some (e.g., Slote, 2001) argue that an act is good as long as the decision maker has good intentions, most scholars agree that the goodness of an act cannot solely be defined by the decision maker's motives. Good intentions need to be paired with good acts (Zagzebski 2004). This approach is more objective and acknowledges that biases may distort a decision maker's motivations and motives. While the goodness of an act can potentially be assessed post hoc, individuals' intentions are impossible to observe. Moreover, individuals are often unable to explain their intentions, making them difficult to measure.

Explainability has been the focus of much AI research and scholars have developed several algorithmic tools to improve the transparency of AI decisions (e.g., Doran et al. 2017; Lundberg and Su-In 2017; Ribeiro, Singh, and Guestrin, 2016; Shrikumar, Greenside, and Kundaje 2017). Our review indicated that Doran's framework has the most potential to serve as the foundation to create a virtue ethics based explicit moral agent since the authors argue that AI systems should be augmented by a reasoning engine that uses one of the previous explainability methods to abstractly mimic how humans extract the critical features for decision making. However, a common property between all current AI methods is that they only consider the explanatory variables during inference. In other words, the explainable deduction are artefacts rather than a core component shaping the way the machine makes decisions. We claim that to successfully create a virtue ethics based explicit moral agent requires that the explainable policies are trained with explainable deductions identified by a previous inference step. Only then can the source of unethical conduct be identified, understood, and corrected. This process is similar to a parent telling a child: "don't do that because you may hurt someone" and not simply "don't do that". This is important since it will allow AI to communicate its intentions. More importantly, when unethical conduct is detected, developers can analyze and correct the source of the conduct, thereby teaching AI to engage in acts a virtues person would engage in. Hence, we extend the framework of Doran et al. (2017) by adding an additional training step after the generation of the explainable deductions. By ensuring that the machine includes the explanatory variables to learn the critical features to make decisions, one can verify whether the learned features are similar to those of a virtuous person, at least from an interpretability perspective.

As described in Figure 1, our modified framework relies on two sequential steps:

1. a regular AI inference phase to generate explainable deductions corresponding to the explainable framework introduced by Doran et al.

2. a contextual AI training phrase that includes the generated explainable deductions as part of the dataset.

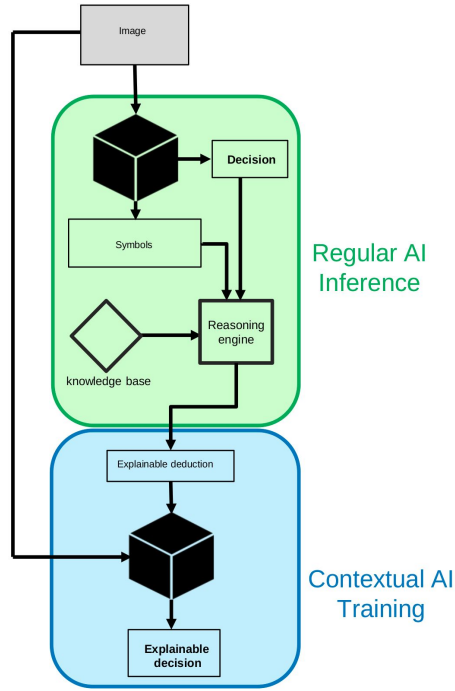

**Figure 1:** Our training framework extending the work of Doran et al. (2017)

While training explainable policies using this framework will take longer, we suggest that such a computational cost is needed not only to make the policy dependent on the explainable variables, but also to have stronger metrics to assess the quality of the learned policy. In fact, the second step introduced in this framework not only contextualizes the output of the policy but also provides a new paradigm for evaluating the AI models. During the testing phase of self-driving cars for example, license providers can fix the input of the model (e.g. image) and only vary its context. This methodology extends the performance assessment biased toward the system's accuracy by evaluating the impact of the explainable variables on the choice of the policy. Furthermore, the explainable deductions can be used as a proof by systems' designers to explain wrong policy decisions of AI systems. Self-driving cars would be able to present the explainable deduction to the police or insurance company to assess the car's reasoning and appropriateness of the policy outcome. If the action taken by the car was in fact inappropriate based on the assessment of the explainable deduction, changes to the domain knowledge base can be made to ensure proper future conduct. Additionally, one can control the values of the explainable variables toward some target values for specific applications to guarantee a desired behavior. In this way, regulators can drive the way the domain knowledge base is shaped by providing specific explainable variables that AI models are expected to use during their training. The explanatory variables can thus bridge the gap between engineering teams and regulatory agencies.

**Discussion and Conclusion**

In line with Anderson and Anderson (2007), most AI scholars follow an outcome or action-driven approach and focus on principles that need to be followed to ensure ethical conduct. However, as discussed, this approach has numerous shortcomings. While some actions may be universally right or wrong, others depend on the context. The Heinz dilemma (Colby et al. 1987) portraits a classic example: Should a husband steal a life saving drug for his dying wife? What if the pharmacist declined a payment plan and reasonable non-monetary compensation? Stealing is wrong but considering the context and the intention, one might reason that it is ok for Heinz to steal the medicine and save his wife's life. Our framework takes contextual information that contains the critical granular features that shape the best action to take into account. Moreover, this information is readily available for audits and can be used to improve future reasoning. Hence, we believe that our framework better represents how individuals make decisions and can be used to develop virtuous machines.